\documentclass[twocolumn]{aastex631}

\usepackage{amsmath}
\usepackage{makecell}
\usepackage{tabularx}
\usepackage{hyperref}

\usepackage{listings}
\usepackage[T1]{fontenc}
\usepackage{booktabs}
\newcommand{\p}{\partial}
\newcommand{\ie}
{\textit{i}.\textit{e}.\hspace{1mm}}

\begin{document}

\title{Meridional circulation molecular-weighted}
\author[0000-0002-7179-8254]{Deepayan Banik}
\affiliation{Department of Physics, University of Toronto,
60 St George Street, Toronto, Ontario, M5S 1A7, Canada}

\author{Kristen Menou}
\affiliation{Department of Physics, University of Toronto,
60 St George Street, Toronto, Ontario, M5S 1A7, Canada}
\affiliation{Dept.   of  Physical  \&  Environmental  Sciences,  University  of  Toronto  Scarborough,\\  1265 Military Trail, Toronto, Ontario, M1C 1A4, Canada}
\affiliation{David A. Dunlap Department  of Astronomy \& Astrophysics, University of Toronto, \\
50 St.  George Street, Toronto, Ontario, M5S 3H4, Canada}

\author{Evan H. Anders}
\affiliation{Kavli Institute for Theoretical Physics, University of California, Santa Barbara, Santa Barbara, CA, 93106, USA}

\begin{abstract}

Meridional circulation in stratified stellar/planetary interiors in the presence of stable molecular-weight gradients remains poorly understood, thereby affecting angular momentum transport in evolutionary models. We extend the \textit{downward control principle} of atmospheric sciences to include compositional stratification. Using a linearized analysis we show that stable $\mu$-gradients slow down the penetration of circulation into the depths, emphasizing the importance of time-dependent solutions. However, additional effects such as horizontal turbulence or magnetic fields are needed to halt it completely. We also find limits demarcating linear and nonlinear regimes in terms of Schmidt and Rossby numbers. Nonlinear simulations exhibit compositional mixing due to meridional currents, enabling deeper penetration than otherwise. We propose slowly evolving and steady-state scenarios for the solar tachocline e and helioseismically observed heavy element abundances, while acknowledging the absence of constraints on the radial variation of the Schmidt number. In the context of stellar evolution, differential rotation profiles of solar-type main-sequence stars may follow analytical solutions extending from \cite{banik2024meridional}, thereby aiding further probes into magneto/hydrodynamic instabilities and outward angular momentum transport.

\end{abstract}




\section{Introduction}


The study of meridional circulation --- a slow, global flow occurring between the poles and equator in low-mass stellar radiative envelopes --- historically lacks consensus in the astrophysics community due to degeneracies in modeling parameters like the background rotation rate, shear turbulence, magnetic fields, or mean-molecular-weight gradients; see the introduction of \cite{zahn1992circulation}. This circulation was first conceptualized by \cite{eddington1925circulating} as a restorative action to centrifugal effects of rotation, resulting in a hotter pole and cooler equator. Later, this baroclinically driven Eddington-Sweet (ES) circulation was found to be insignificant \citep{busse1981eddington} and the spin-down effect caused by escaping stellar winds \citep{howard1967solar,greenspan1963time}, was made responsible for a differential rotation that prevailed through the overlying convection zone to resistively torque the faster rotating radiative zone underneath giving rise to a stronger meridional circulation \citep{goode1991we,zahn1992circulation}. {\color{black}In the solar context, the radiatively driven hydrodynamic penetration model of \cite{spiegel1992solar} has stood the test of time. However, their assumptions of  anisotropic diffusion coefficients for `tachocline' confinement has seen recent challenge following a better understanding of stratified shear turbulence that occurs at this famous radiative-convective boundary \citep{garaud2025towards}. }

\begin{figure}
    \centering
    \includegraphics[width=0.8\linewidth]{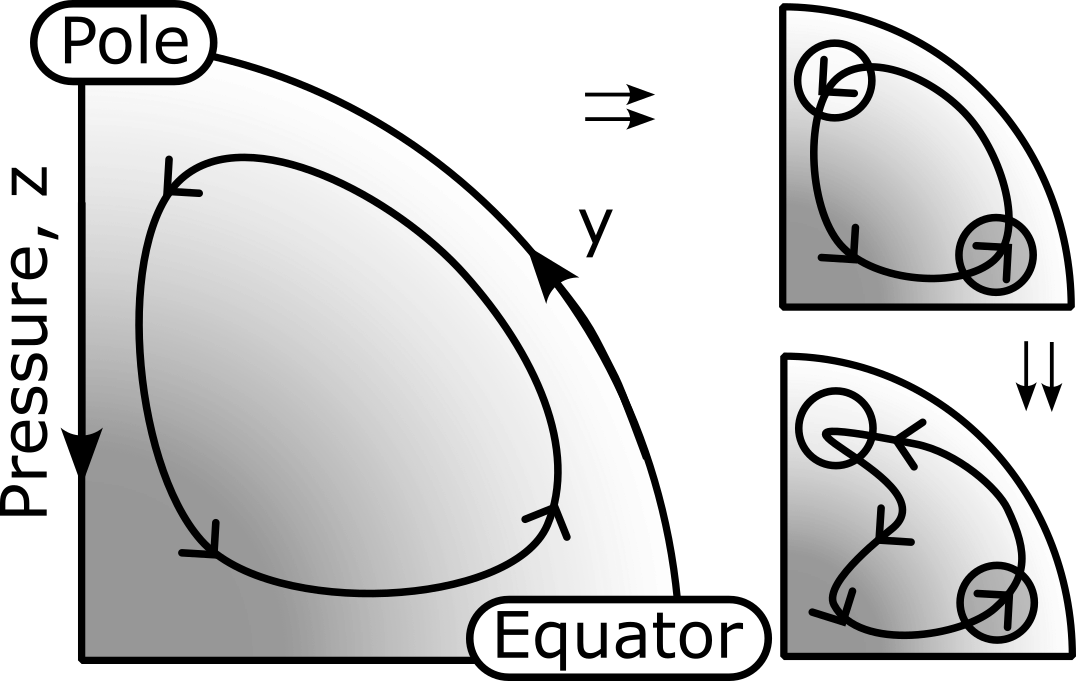}
    \caption{Schematic of meridional circulation in the presence of $\mu$-gradients. Initially, the background gradient of mean molecular weight, shown as a grey-white shade with grey denoting heavier material, is simply advected by the rotational meridional current. The resulting $\mu$ distribution (white and grey circles) reacts back on the circulation, affecting it in a way described by the thermal wind equation (equation \ref{eq:20}). The causal pathway is described in \citealp{mestel1953rotation}.}
    \label{fig:1}
\end{figure}

It is also known that radiative zones in stars develop molecular-weight($\mu$) gradients over their nuclear lifetimes due to core-seated nuclear reactions or gravitational settling of Helium from the outer convective envelope \citep{stringfellow1983evolutionary,proffitt1991gravitational}, thus spanning the entire radiative zone. The role of such gradients in determining the fate of meridional circulation is still highly debated. \cite{mestel1953rotation} was the first to address this problem. He argued that the ES circulation leads to a non-spherical distribution of $\mu$ in compositionally stratified regions, resulting in $\mu$-currents that oppose the initial circulation, thereby truncating the radiatively rendered laminar penetration predicted in \cite{sweet1950importance}; see Figure \ref{fig:1}. Aside from the transport of angular momentum, this theory was also supported in the context of compositional mixing \citep{kippenhahn1974circulation} and used to explain abundances of Am stars \citep{vauclair1977time}. On the contrary, starting from the zonal torquing scenario, \cite{spiegel1992solar} showed that a stable $\mu$-gradient produces a diffusive term that adds to the laminar penetration of meridional circulation, effectively contradicting the choking argument of \cite{mestel1953rotation}. Additionally, anisotropic turbulence was invoked, not only to justify choking of meridional circulation near the \textit{tachocline}, but also to smooth out the horizontal compositional inhomogeneities made responsible for the generation of $\mu$-currents earlier, thus mitigating their effect on the said circulation \citep{zahn1992circulation}. Despite all efforts, the inherent nonlinearity of the problem has only been alluded to but has not been self-consistently solved so far \citep{mestel1953rotation,huppert1977penetration}. 

Stellar evolution models like MESA \citep{paxton2010modules} use diffusion coefficients to parameterize rotationally induced mixing caused by hydrodynamic instabilities and meridional circulation \citep{heger2000presupernova}. For the latter, the less relevant ES circulation is considered. Additionally, it is assumed to be stabilized by $\mu$-gradients according to \cite{mestel1953rotation}. The improvement offered by spin-down theory and the contradictory effect of $\mu$-gradients is noted in \cite{heger2000presupernova}, but the corresponding fourth-order differential equation for differential rotation is not solved exactly owing to increased computational effort. Rather, this and other complications are dealt with by empirically calibrating diffusion coefficients based on the observed surface abundances of stars.


Newer numerical studies on meridional circulation use a more complete description of the problem, including the impact of interfacial conditions, radiative-convective coupling, magnetic fields, and complete three-dimensionality \citep{garaud2008penetration,garaud2009penetration,wood2011sun,wood2012transport,matilsky2022confinement,korre2024penetration,matilsky2024solar}. However, $\mu$-gradients are rarely considered. Those that do, focus on instabilities like overturning double-diffusive convection and fingering convection \citep{traxler2011numerically,rosenblum2011turbulent,garaud2021double}. \cite{wood2011polar} discusses the role of compositional gradients in confining the polar magnetic fields. {\color{black}A series of three papers have discussed the interaction of $\mu$-gradients created by Helium settling and rotational mixing due to meridional circulation \citep{vauclair2003coupling, theado2003coupling, paper3}. However, in an attempt to include multiple physical processes, like horizontal turbulence, Eddington-Sweet flows, differential rotation and $\mu$-gradients, major approximations are used, and their conclusions focus more on application to stellar populations than the self-contained fluid dynamics problem. Moreover, similar to prior work on particle transport and rotational mixing in stars, the effects of inherent nonlinearities are not explored \citep{charbonneau1988two, charbonneau1991meridional}. Recognizing such limitations and empirical parameterizations in stellar evolution models, it is thus crucial to extend our understanding of meridional circulation under the influence of compositional stratification. }

Recently, \cite{banik2024meridional} demonstrated the use of the `stratospheric downward control principle' \citep{haynes1991downward} to explain stellar meridional circulation from an atmospheric science perspective; see also Ch. 8 of \cite{hughes2007solar} for a description of Haynes-Spiegel-Zahn burrowing. Under the simplifying assumptions of axisymmetry, f-plane, linear relaxation, and thermal wind balance, the model reproduced the results of meridional circulation in stellar physics, offering analytical solutions to a second-order advection-diffusion equation as opposed to the fourth-order equation presented in \cite{spiegel1992solar}. In this paper, we extend this model by incorporating $\mu$-gradients into the formalism. We restrict ourselves to non-magnetic stars to simplify the dynamics of our system while acknowledging their possible role in restricting the spread of the solar tachocline \citep{gough1998inevitability,brun2006magnetic,strugarek2011magnetic,matilsky2022confinement}. Stratification regimes stable according to both the Schwarzchild and Ledoux criteria are studied to help moderate the debate of \cite{mestel1953rotation} and \cite{spiegel1992solar}.

{\color{black}In \S \ref{sec:2} we derive a linearized 1D inhomogeneous diffusion equation for the zonal wind akin to \cite{banik2024meridional}, incorporating molecular-weight stratification. In \S \ref{sec:3}, we derive dimensionless governing equations and identify regimes of the parameter space where nonlinear effects dominate, in terms of Prandtl, Schmidt, and Rossby numbers, and make comments about the Sun. We present numerical simulations in with Jupiter-like parameters in Dedalus2 in \S \ref{sec:4}, first reproducing the solutions of \cite{showman2006deep} and then extending them to cover effects of $\mu$-gradients and nonlinearities, discussing their impact on the meridional circulation. In \S \ref{sec:5} we delineate the implications of our results in the context of angular momentum evolution of main-sequence stars and propose tentative pictures of the solar tachocline. Finally, we conclude in \S \ref{sec:6}.}




\section{1-D linearized equations} \label{sec:2}

The treatment presented in this section is similar to \cite{banik2024meridional} with all assumptions applicable. We re-describe the setup here, however, the reader is directed to the corresponding paper for details and clarification. The vertical and meridional coordinates are denoted by pressure $p$ and distance $y$, respectively, with time represented by $t$. The equations governing zonal-mean momentum, mass continuity, and thermodynamic energy conservation for an axisymmetric stratified hydrostatically balanced fluid on an f-plane, subjected to momentum forcing at the top are,

\begin{equation}
    \frac{\partial u}{\partial t} - f v = \frac{u_F - u}{\tau_F},
    \label{eq:1}
\end{equation}

\begin{equation}
    \frac{\partial v}{\partial y} + \frac{\partial w}{\partial p} = 0, \hspace{1mm} \text{and}
    \label{eq:2}
\end{equation}

\begin{equation}
    \frac{\partial T'}{\partial t} - \left( \frac{N^2 H^2}{\mathcal{R}} \right) \frac{w}{p} = \frac{T'_E - T'}{\tau_R}
    \label{eq:3}
\end{equation}

Here, $u$, $v$, and $w (\equiv \partial p / \partial t)$ represent the zonal, meridional, and vertical wind speeds in the rotating frame. The Coriolis parameter $f (= 2 \Omega \sin \theta)$ is a constant that depends on the specific f-plane latitude $\theta_o$ chosen. $\mathcal{R}$ is the specific gas constant. Momentum forcing relaxes $u$ towards $u_F$, and thermal forcing relaxes $T$ towards $T'_E$ over frictional drag ($\tau_F$) and radiative ($\tau_R$) timescales, respectively. The Brunt frequency $N$ for a stratified radiative zone, with a temperature profile $T = T_o(\ln p) + T'(y, \ln p, t)$, where $T'$ is a perturbation from the base profile $T_o$ ($= T_{\text{min}} + T_{o,\ln p} \cdot \ln p$), is defined by

\begin{equation}
    N = \sqrt{ \frac{g}{H} \left( \frac{\mathcal{R}}{C_p} - \frac{\mathrm{d} \ln T_o}{\mathrm{d} \ln p} \right) },
    \label{eq:4}
\end{equation}

where $C_p$ is the specific heat at constant pressure, $H = \mathcal{R}T_{o}/g$ is the isothermal scale height, and $g$ is the acceleration due to gravity. {\color{black} Despite their apparent pressure dependence, we take $C_p, H$ and $g$ to be constant through the radiative zone, a standard assumption pertaining to the Anelastic approximation in atmospheric physics \citep{holton1973introduction, showman2006deep}. }

We now incorporate the effects of molecular-weight$\mu$. Similar to temperature, the quantity is split into a background $\mu_o$ and a perturbation $\mu'$ such that $\mu = \mu_o(\ln p) + \mu'(y, \ln p, t)$. Further, $\mu_o$ is subdivided into a constant $\mu_{\text{min}}$ and a pressure dependent component, $\alpha\ln p$, where $\alpha(\equiv \p \mu_o / \p\ln p)$ represents the log-pressure rate of change of $\mu_o$ \ie $\mu_o = \mu_{\text{min}} + \alpha\ln p$. We use a linear relaxation to replace diffusion, similar to temperature and zonal velocity. Accounting for axisymmetry, the full nonlinear $\mu$ conservation equation is as follows,
\begin{equation}
    \frac{\p \mu}{\p t} +v \frac{\p \mu}{\p y}+ \frac{w}{p}\frac{\p\mu}{\p\ln p} = \frac{\mu_{o}-\mu}{\tau_{\mu}}
    \label{eq:5}
\end{equation}
{\color{black} Here, $\tau_{\mu}$ might, in principle, represent the nuclear time for a star, or an appropriate timescale for a planet}, implying that the flow-advected molecular-weightdistribution ($\mu$) is linearly relaxed to the background profile $\mu_o$ over this time. Substituting $\mu$ with $\mu_o + \mu'$, we get the evolution equation for the $\mu$-perturbation ($\mu'$),
\begin{equation}
    \frac{\p \mu'}{\p t} + \frac{w}{p} \alpha = -v\frac{\p\mu'}{\p y} -\frac{w}{p}\frac{\p\mu'}{\p\ln p} - \frac{\mu'}{\tau_{\mu}}
    \label{eq:6}
\end{equation}

{\color{black}The second enhancement to the system comes through the thermal wind balance, a combination of geostrophic balance and hydrostatic balance, that takes a more involved form as compared to \cite{banik2024meridional}. The gas constant $\mathcal{R}(=\mathcal{R}_u/\mu)$ now depends on the state variable $\mu$. Deriving from first principles (see Appendix \ref{app:tw}), the compositionally modified thermal wind relation becomes,
\begin{equation}
f p\frac{\p u}{\p p} = g H\frac{\p }{\p y}\left(\frac{T'}{T_o} - \frac{\mu'}{\mu_o}\right)
    \label{eq:8}
\end{equation}
where $T_o$ and $\mu_o$ are mean background quantities. This allows us to account for the contribution of horizontal gradients of molecular-weightto the vertical zonal wind shear, an indicator of penetration of the wind. Since temperature decreases and mol-wt increases with pressure for stable stratification, their flow advected meridional gradients would be of opposite signs as also seen in simulations later. Thus, both terms in equation \eqref{eq:8} contribute to increasing the vertical shear of the zonal wind, consequently resisting penetration. }

We consider external thermal/baroclinic forcing in equation \eqref{eq:3} to be zero \ie $T'_E=0$ to focus on the role of momentum forcing ($u_F$) on meridional circulation. We adopt the two-layer setup described in \cite{banik2024meridional}. The top layer experiences zonal momentum forcing through a non-zero $u_F$ while the bottom layer is passive ($u_F=0$).

The first two terms on the right-hand side of equation \eqref{eq:6} are nonlinear and will be ignored as a simplifying assumption in our preliminary analytical work.  Using equations \eqref{eq:1}-\eqref{eq:3}, \eqref{eq:6}, and \eqref{eq:8}, we derive a single equation for the time-dependent penetration of the zonal wind in the presence of a background molecular-weight gradient. Assuming the variables $u$, $v$, $w$, $T'$ and $\mu'$ to be meridionally periodic with a wavenumber $l$, so that
\begin{equation}
\begin{split}
    u=\bar{u}\cos(ly), v=\bar{v}\cos(ly), w=\bar{w}\sin(ly), \\T'=\bar{T}\sin(ly), \text{and} \hspace{1mm} \mu'=\bar{\mu}\sin(ly)
\end{split}
    \label{eq:9}
\end{equation}
where the barred quantities are functions of $t$ and $p$, we obtain the following forms of equations \eqref{eq:1}, \eqref{eq:2}, and \eqref{eq:8} in the passive bottom-layer below the momentum-driven top-layer (cf. equation (7) and related geometry in Figure 2 of \citealp{banik2024meridional}, and Figure \ref{fig:2} of this paper),

{\color{black}
\begin{equation}
\begin{split}
    \frac{\p \bar{u}}{\p t}-f \bar{v}=\frac{-\bar{u}}{\tau_F}, \hspace{5mm} l\bar{v}=\frac{\p \bar{w}}{\p p}, 
    \hspace{1mm} \\ \text{and} \hspace{1mm} f\frac{\p \bar{u}}{\p \ln p} = gHl \left(\frac{T'}{T_o} - \frac{\mu'}{\mu_o}\right).
    \end{split}
    \label{eq:10}
\end{equation}
Additionally, assuming rapid thermal and molecular-weightadjustments  ($\p T'/\p t = 0$, $\p \mu'/\p t = 0$), equations \eqref{eq:3} and \eqref{eq:6} give,
\begin{equation}
   \bar{T} = \left(\frac{N^2 H^2 \tau_R}{\mathcal{R}}\right) \frac{\bar{w}}{p}, \hspace{2mm}\text{and} \hspace{2mm} \bar{\mu} = - (\alpha \tau_{\mu}) \frac{\bar{w}}{p}.
    \label{eq:11}
\end{equation}
Combining equations (\ref{eq:10})-(\ref{eq:11}) gives us a second-order inhomogeneous diffusion equation for the zonal wind in log-pressure ($z= H \ln p$) coordinates:

\begin{equation}
\begin{split}
    \frac{\p \bar{u}}{\p t} + \frac{\bar{u}}{\tau_F} = \frac{1}{Q} \left(H\frac{\p u}{\p z} + H^2\frac{\p ^2 u}{\p z^2}\right);
    \\ \text{where,} \hspace{1mm} Q = \frac{gH}{f^2/l^2}\left(\frac{N^2H^2 \tau_R}{R T_o} + \frac{\alpha \tau_{\mu}}{\mu_o}\right). 
\end{split}
    \label{eq:12.1}
\end{equation}
We thus have a 1D linear partial differential equation for the evolution of zonal wind in the presence of a background gradient of molecular-weight$\alpha$. In its most general form, the quantity $Q$ with units of time may be a function of the vertical coordinate taking the same form as equation (16) in \cite{banik2024meridional}, which may be recovered by setting $\alpha=0$. However, for this work, we consider it to be constant through the radiative zone. In fact, it is important to clarify here that the pressure dependence of $T_o$ and $\mu_o$ only features in the Brunt frequency, $N$, and the background mol-wt gradient, $\alpha$, in equations \eqref{eq:3} and \eqref{eq:6}. For the rest of the occurrences, they may be replaced by $T_{\text{min}}$ and $\mu_\text{min}$, respectively. This is because the variations due to stratification of temperature and mol-wt are assumed to be smaller than background quantities, similar to \cite{showman2006deep}. Note here that the zonal wind is intrinsically connected to the meridional circulation \citep{banik2024meridional}. Therefore, the degree of penetration of the zonal wind and meridional circulation may be considered equivalent, and hence will be used interchangeably in this context for the rest of the paper.

\hspace{-0.75cm}
\begin{figure*}
    \centering
    \includegraphics[width=0.8\linewidth]{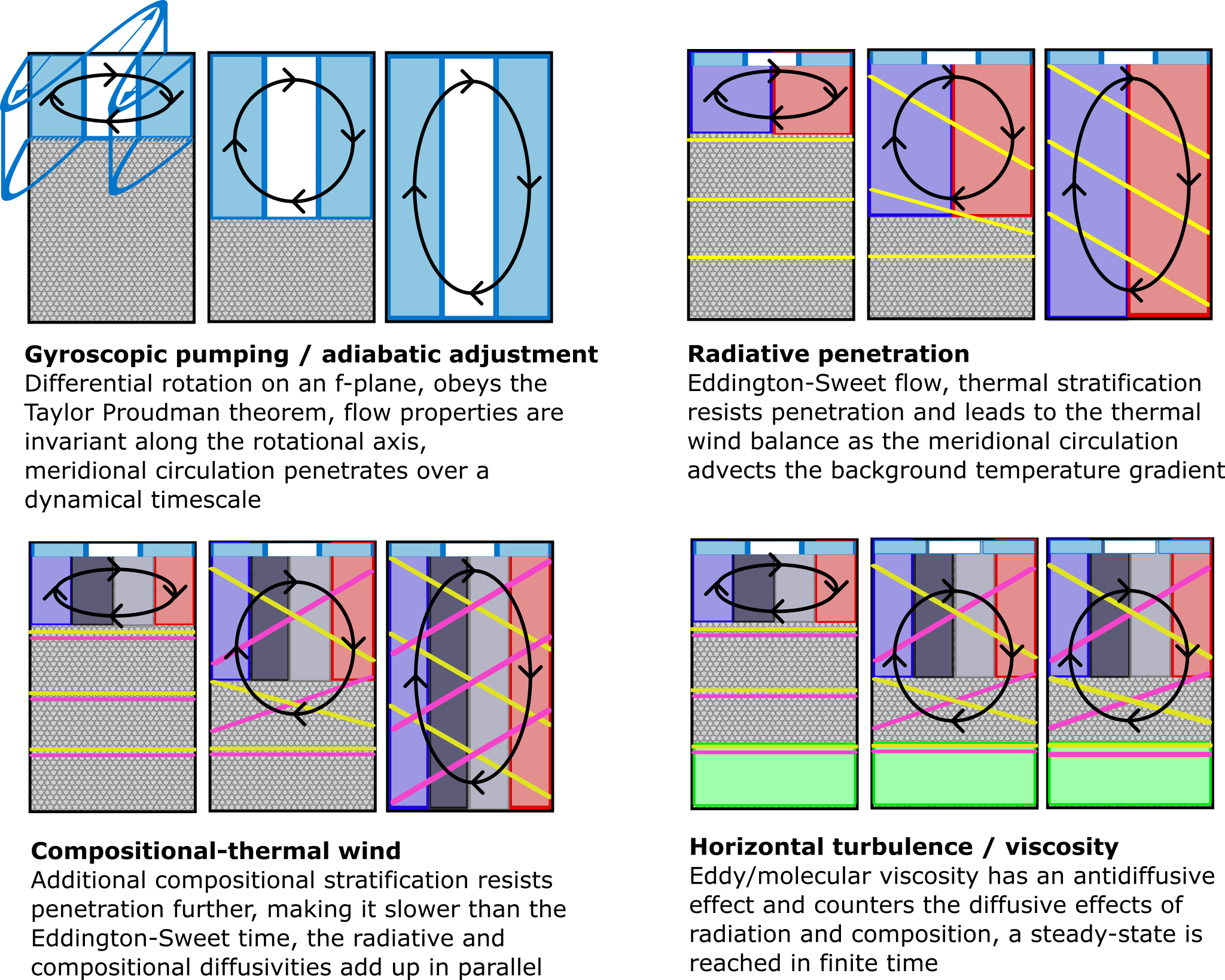}
    \caption{The process of penetration of zonal forcing (top-left panel) and corresponding meridional circulation (circular lines with arrows) under effects of pure rotation, thermal stratification, compositional stratification, and horizontal momentum transport via. eddy/molecular viscosity. The individual panels resemble f-planes similar to the computational domain shown later in Figure \ref{fig:2}. (Top left) Gyroscopic pumping: Blue and white regions correspond to regions of opposing zonal velocity. The zonal velocity and meridional circulation penetrate together all the way to the bottom of the domain following the Taylor Proudman theorem over a dynamical timescale \citep{showman2006deep}. (Top right) Radiative penetration: The yellow isotherms from a stable thermal stratification are tilted by the adiabatic meridional circulation, thereby resisting the penetration, which eventually proceeds slowly due to radiative diffusion over the Eddington-Sweet timescale \citep{spiegel1992solar}. (Bottom left) Compositional-thermal wind: The magenta lines of constant mean-molecular weight from compositional stratification are tilted in the opposite direction by the meridional flow, additionally resisting it and making the penetration even slower, yet reaching the bottom at steady-state. (Bottom right) Horizontal turbulence/viscosity: Horizontal momentum transport by eddy viscosity or magnetic fields oppose the penetration, thereby halting it at a finite depth and rendering a faster steady-state \citep{spiegel1992solar,gough1998inevitability}. Red and blue regions represent fluid of relatively higher and lower temperatures, and dark grey and light grey regions represent higher and lower mean-molecular weight, respectively. The green region is an approximate representation of the opposition offered by viscous or magnetic effects.}
    \label{fig:1.5}
\end{figure*}

Equation \eqref{eq:12.1} shows that the presence of a stable compositional stratification reduces the net diffusion coefficient $1/Q$, thus making the penetration of zonal wind slower, as shown in \cite{mestel1953rotation}. Through elementary algebraic manipulation it may also be shown that the thermal and compositional diffusion coefficients add up in parallel, making the effective diffusion coefficient smaller than both. Consequently, the inclusion of a compositional stratification increases the effective Eddington-Sweet time. However, the burrowing of the tachocline will not be halted unless a resisting mechanism such as horizontal drag acts on the system \citep{spiegel1992solar}. See Figure \ref{fig:1.5} for a complete description of the process.  The above conclusion contradicts those in Appendix B of \cite{spiegel1992solar}, which say that compositional gradients aid the diffusion of differential rotation through the depths of radiative zones. We have thus re-established the conclusions of \cite{mestel1953rotation} for a zonally-forced radiative interior, in addition to the regular Eddington-Sweet currents originating from rotational flattening, as was the setup in his paper. 

Taking a ratio of the second to the first term in the expression for $Q$, we may derive a limit at which mol-wt contributions to the flow are insignificant. This is given by,
\begin{equation}
    \frac{ \tau_{\mu}}{\tau_R} \cdot \frac{{\bar{\alpha}}gH}{N^2H^2} \ll 1.
    \label{eq:11.5}
\end{equation}
where $\bar{\alpha}= \alpha/\mu_o$. The same limit may be obtained from the scaled thermal wind relation later in equation \eqref{eq:19} by taking ratios of the terms on the RHS.

Analytical solutions to equation \eqref{eq:12.1} may be obtained similar to \cite{banik2024meridional}, acting as improved prescriptions for differential rotation profiles in evolutionary codes because of the additional physics of compositional stratification. 
}

\section{2-D nonlinear dimensionless equations}\label{sec:3}

We now go back to our original nonlinear equations and simplify them by introducing a new variable for the vertical velocity $(\Tilde{w})$, such that, $$\frac{w}{p} = \frac{d \ln p}{dt} = \frac{1}{H}\frac{dz}{dt}=\frac{\Tilde{w}}{H}.$$ The convention in atmospheric sciences is to use $dz = -Hd\ln p$ \citep{holton1973introduction}, indicating that $z$ decreases as pressure increases. We drop the negative sign to cater to stellar astrophysics contexts by having our vertical coordinate increase with pressure as one moves from the top of the radiative zone to the bottom, as in \cite{banik2024meridional}, making $\alpha$ positive contrary to the convention in Appendix B of \cite{spiegel1992solar}. The $z$-dependent formulation helps us eliminate domain-dependent non-constant coefficients beneficial for the numerical implementation in the spectral solver, Dedalus2, described later. Thus, in the mass balance equation \eqref{eq:2}, we have, $$\frac{\p w}{\p p} = \frac{\Tilde{w}}{H}+\frac{\p \Tilde{w}}{\p z}.$$

We now nondimensionalize our system to facilitate general applicability. The following parameters are used for nondimensionalization:
\begin{equation}
\begin{split}
\{z,y,t,\tau_F,\tau_R,\tau_{\mu}\} = \{H z^*, L_y y^*, \tau_o( t^*,\tau_F^*,\tau_R^*,\tau_{\mu}^*)\} \\
    \{u,v,\Tilde{w},T',\mu'\}=\{U_F u^*,Vv^*,W\Tilde{w}^*,T_{s}T^*, \mu_{s}\mu^*\}\\    
\end{split}\label{eq:14}
\end{equation}
Here, $T_s$ and $\mu_s$ are representative quantities determined from an order-of-magnitude analysis later in \S \ref{sec:3.1}, and {\color{black}$\tau_o$ may be chosen arbitrarily to be one of $\tau_{\mu}, \tau_F$ or $\tau_R$}.

Scaling equations \eqref{eq:1}, \eqref{eq:2}, and \eqref{eq:3} using the above, and adding nonlinear terms to the energy equation, we get,
\begin{equation}
    \frac{\p u^*}{\p t^*}- v^* \frac{f(\tau_o V)}{U_F}=\frac{u_F^*-u^*}{\tau_F^*} = \frac{\tau_o}{\tau_F}(u_F^*-u^*),
    \label{eq:15}
\end{equation}

\begin{equation}
    \frac{\p v^*}{\p y^*}+\left[\Tilde{w}^* + \frac{\p \Tilde{w}^*}{\p z^*}\right] \left(\frac{W}{H}\cdot\frac{L_y}{V}\right)= 0, \label{eq:16}
\end{equation}

\begin{equation}
\begin{split}
   \frac{\p T^*}{\p t^*} - \Tilde{w}^* \frac{N^2 H^2}{\mathcal{R}H} \cdot \frac{(\tau_o W)}{T_{s}} = \hspace{4cm}\\ -\Tilde{w}^* \frac{\p T^*}{\p z^*}\frac{(\tau_o W)}{H} - v^*\frac{\p T^*}{\p y^*}\frac{(\tau_o V)}{L_y} 
   - \frac{\tau_o}{\tau_R}T^*.
\end{split}
 \label{eq:16}
\end{equation}

The scaled molecular-weightconservation equation \eqref{eq:6} is, 
\begin{equation}
\begin{split}
\frac{\p \mu^*}{\p t^*} +  \Tilde{w}^* \frac{\alpha}{H} \cdot\frac{(\tau_o W)}{\mu_{s}}=\hspace{5cm}\\-\Tilde{w}^* \frac{\p \mu^*}{\p z^*}\frac{(\tau_o W)}{H} - v^*\frac{\p \mu^*}{\p y^*}\frac{(\tau_o V)}{L_y} -\frac{\tau_o}{\tau_{\mu}}\mu^* \\
\text{or,} \hspace{2mm}\frac{\p \mu^*}{\p t^*}+  \Tilde{w}^* \frac{\alpha}{H} \cdot\frac{(\tau_o W)}{\mu_{s}}= -\frac{\tau_o}{\tau_{\mu}}\mu^*\hspace{1mm} \text{in the linear limit,}
\end{split} \label{eq:17}
\end{equation}
and the thermal wind relation \eqref{eq:8} becomes,


{\color{black}
\begin{equation}
    \begin{split}
        \frac{\p u^*}{\p z^*}=\frac{\mathcal{R}T_{s}}{fU_F L_y}\left[\frac{\p T^*}{\p y^*}-\left(\frac{\mu_{s}}{\mu_o}\cdot\frac{T_o}{T_s}\right) \frac{\p \mu^*}{\p y^*}\right].
    \end{split} \label{eq:19}
\end{equation}
}

\subsection{Linear order-of-magnitude scaling}\label{sec:3.1}

A leading order linearized balance at steady state in equations \eqref{eq:15}-\eqref{eq:19}  yields approximate magnitudes of $V, W, T_{s}$  and $\mu_{s}$ in terms of the known parameters $f,N,H, U_F, L_y, \mathcal{R}_u, \alpha, \mu_{\text{min}}, \tau_F, \tau_R$, and $\tau_{\mu}$ as follows,
{\color{black}
\begin{equation}
\begin{split}
    V\sim\frac{U_F}{f \tau_F}, \hspace{1mm} W \sim\frac{V}{L_y}H, \hspace{1mm} T_{s} \sim \frac{N^2 H^2 \mu_{\text{min}}}{\mathcal{R}_u H} \tau_R W \\ \hspace{1mm} \text{and,} \hspace{1mm} \mu_{s}\sim \frac{\alpha}{H} \tau_{\mu} W.
\end{split}
    \label{eq:20}
\end{equation}
}
{\color{black} The scale $V$ matches that derived in equation (31) of \cite{garaud2025towards} where $r_{cz}^2/\nu_h$ is equivalent to our $\tau_F$, with $r_{cz}$ being a characteristic length scale of the flow and $\nu_h$ is the horizontal eddy viscosity. 

Note here that the above scalings apply only at steady state. Also, $V$ may represent the meridional velocity at any depth depending on the local value of $\tau_F$. Close to the convection zone $\tau_F$ is small compared to the depths, and hence $V$ is larger. \cite{garaud2025towards} elucidates one such scenario where horizontal diffusion coefficients are dependent on the local amplitude of stably stratified shear. $V$, and consequently all other quantities also depend on the local scaling for zonal velocity ($U_F$ above) which in turn depends on the degree of penetration through stratification. An extension of this discussion for different degrees of penetration, and the corresponding appropriate scaling for zonal velocity is provided in Appendix \ref{app:scale}.}

\subsection{Limits on linearity}\label{sec:3.2}

We drop the asterisk while referring to nondimensional quantities for ease of writing. The first two terms on the RHS of equation \eqref{eq:17} are nonlinear, and shall only affect the perturbation $\mu$ field if significant. The linear limit occurs when the coefficient of the nonlinear terms is much smaller than the linear terms. Then, we have,
{\color{black}
\begin{equation}
    \frac{\alpha W}{H}\cdot \frac{\tau_o}{\mu_s} \gg \frac{W}{H}\tau_o \hspace{1cm} \text{or} \hspace{1cm} \frac{\alpha}{\mu_s} \gg 1.
 \notag
\end{equation}
}

Substituting for relevant linear limit quantities from equation \eqref{eq:20}, this is simplified to,
\begin{equation}
    \frac{\tau_{\mu}}{\tau_F}\cdot\frac{U_F}{f L_y} \ll 1, \hspace{1cm} \text{or} \hspace{1cm} Sc\cdot Ro\ll 1\label{eq:21}
\end{equation}
where $Sc$ is the Schmidt number describing the ratio of momentum and mean-molecular-weight diffusion, and $Ro$ is the Rossby number of the system. {\color{black} We find that the Rossby number of the system may be defined in multiple ways owing to different scalings of zonal and meridional velocities $u$ and $v$. The conclusions of this work are, however, largely agnostic to the alternative forms of the number. A discussion on the same is included in Appendix \ref{app:rossby}. The limits expressed above and the effect of nonlinearities are explored later in \S \ref{sec:4}.}

As an aside, one might reproduce the conditions under which temperature nonlinearities are insignificant,
\begin{equation}
    \frac{\tau_{R}}{\tau_F}\cdot\frac{U_F}{f L_y} \ll 1, \hspace{1cm} \text{or} \hspace{1cm} Pr\cdot Ro \ll 1\label{eq:22}
\end{equation}
The impact of temperature nonlinearities has not been explored in this paper.

{\color{black}
\subsection{Relevance for Sun-like stars}

From the discussion above, we get two limits set by inequalities \eqref{eq:11.5} and \eqref{eq:21} that concern the presence of mean-molecular-weight gradients. The first demarcates the regime where $\mu$-gradients are insignificant for meridional circulation. Note that inequality \eqref{eq:11.5} does not hold when the governing equation for $\mu$, equation \eqref{eq:17}, is nonlinear. The second, inequality \eqref{eq:21}, indicates the regime where nonlinear advection of $\mu$ dominates. As we shall see later in section \ref{sec:4}, nonlinear effects erase compositional gradients due to mixing by strong meridional currents, causing a reduced impact on the truncation of meridional circulation compared to the linear case. Combining the above limits, we obtain upper and lower bounds for $\tau_{\mu}$ such that the system experiences maximum impact of $\mu$-gradients on the penetration of meridional circulation. {\color{black} These are given by,

\begin{equation}
\begin{split}
        \left(\frac{N^2 H^2}{\bar{\alpha} g H} \right) \tau_R  \ll \tau_{\mu} \ll \tau_F \left(\frac{1}{Ro}\right) = L_y/V\\ 
        \equiv \tau_{\mu\_ \rm min} \ll \tau_{\mu} \ll \tau_{\rm mer\_adv},
\end{split}
    \label{eq:23.5}
\end{equation}
where $\tau_{\mu\_\rm min}$ corresponds to the limit of fastest diffusion of $\mu$ that allows formation of meridional gradients strong enough to affect penetration for given background thermal and compositional stratification, and radiative timescale. $\tau_{\rm mer\_adv}$ is the timescale for meridional advection, and corresponds to the slowest ${\mu}$ diffusion that prevents mixing due to meridional flows. Note that both timescales depend on a complimentary set of system parameters and are thus independent of each other.}

\begin{table}[ht]
    \centering
    \caption{Independent and derived parameters for the solar radiative zone with rigid boundary (\citealp{kippenhahn1990stellar, banik2024meridional}). The bold quantities represent convective modifications to the top boundary \citep{bretherton1968effect}.}
    \label{tab:solar_radiative_zone}
    \begin{tabular}{lc}
        \toprule 
        Parameter (Symbol) & Value \\
        \midrule \midrule
        Radiative timescale ($\tau_R$) & $3 \times 10^{7}$ yrs \\
        Frictional timescale ($\tau_F$) & $10^{13}$ yrs,  \textbf{1 mth} \\
        Zonal forcing wind speed ($U_F$) & $10$ cm/s \\
        Coriolis parameter ($f$) & $5.4 \times 10^{-6}$ rad/s \\
        Scale height ($H$) & $7 \times 10^{9}$ cm \\
        Buoyancy frequency ($N$) & $2.5 \times 10^{-3}$ rad/s \\
        $\mu$-gradient ($\bar{\alpha} =\partial \ln \mu/\partial \ln p$) & $4.16 \times 10^{-3}$ \\
        Gravitational acceleration ($g$) & $13000$ cm/s$^2$ \\
        Meridional length scale ($L_y$) & $7 \times 10^{10}$ cm \\
        \midrule
        Meridional velocity scale ($V$) & $10^{-15}$, \textbf{0.69 cm/s} \\
        Schmidt no. $\times$ Rossby no. ($Sc\cdot Ro$) & $ 10^{-5}$, $\mathbf{10^{9}}$ \\
        Minimum $\tau_{\mu}$ for impact ($\tau_{\mu\_\rm min}$) & $7.6 \times 10^{17}$ s \\
        Meridional advection time ($\tau_{\rm mer\_adv}$) & $10^{25}, \mathbf{10^{11}}$ \textbf{s} \\
        \midrule
    \end{tabular}
\end{table}

For values corresponding to the solar radiative zone in Table \ref{tab:solar_radiative_zone}, we find the lower and upper bounds to be $O(10^{17})$ s, and  $O(10^{25})$ s, respectively, implying that there must exist a range of $\tau_{\mu}$ spanning eight orders of magnitude for which background $\mu$-gradients significantly affect the penetration of the zonal velocity. {\color{black} Here, we must recognize that the scaling in equation \eqref{eq:20} grossly underestimates the meridional velocities near the solar tachocline, making such flows inconsequential for evolution. In reality, the presence of the turbulent convection zone must lead to a smaller value of $\tau_F$ ($\sim$ 1 month, the eddy turnover time), increasing $V$ by approximately $O(10^{14})$ according to equation \eqref{eq:20}, thereby differing significantly from the case with a rigid boundary above \citep{bretherton1968effect, banik2024meridional}. The stronger meridional flow makes the system nonlinear, causing mixing as shown in the following section. However, the actual scale for $V$ would not be as high as it depends on other properties of the convection zone and stratification of the radiative zone (Appendix \ref{app:scale}); see also equation (69) \cite{wood2011sun} for a better estimate of meridional velocities.  In conclusion, a smaller $\tau_F$ near the tachocline brings down $\tau_{\rm mer\_adv}$ in equation \eqref{eq:23.5} making it more likely for the solar tachocline to enter the nonlinear regime of compositional adjustment.}

\begin{table*}[ht]
  \hspace{-1.75cm}
  \caption{List of simulations in different regimes of nonlinearity discussed in this paper.}
  \label{tab:onlytable}
  \begin{tabular}{@{} l  l  c  c  c  c  c @{}}
    \toprule
    Name 
      & Description 
      & $\tau_{\mu}$ (yr) 
      & $\tau_{R}$ (yr) 
      & $\alpha$ (g/mol) 
      & $Sc \cdot Ro$ 
      & $Pr \cdot Ro$ \\
    \midrule \midrule
    T0  
      & No $\mu$ gradients \citep{showman2006deep} 
      & --- 
      & $10^{4}$ 
      & --- 
      & --- 
      & 4.4 \\
    T1  
      & Nonlinear limit, linear terms 
      & $10^{4}$ 
      & $10^{4}$ 
      & 0.01 
      & 4.4 
      & 4.4 \\
    T2 
      & Nonlinear limit, nonlinear terms 
      & $10^{4}$ 
      & $10^{4}$ 
      & 0.01 
      & 4.4 
      & 4.4 \\
    \bottomrule
  \end{tabular}
\end{table*}

The appropriate value of $\tau_{\mu}$ in the solar interior is uncertain. Several processes could set this value depending on the exact location inside the Sun. By default, the parameter could represent molecular/microscopic diffusion of $\mu$ in linearized form. In that role, it acts approximately as fast as viscosity ($\tau_{\mu} \sim \tau_F \sim 10^{13}$ yrs; \citealp{menou2004local}). In other works, microscopic diffusion of Helium has been calculated to be of the order of $10^9$ yrs \citep{theado2003coupling}. It could also be the timescale on which nuclear processes in the star set the background profile of mean molecular weight; $\sim 10^{10}$ yrs for a star like the Sun. However, this is only relevant close to the core where meridional currents are absent. Moreover, this process acts like a source term rather than a linear relaxation. Finally, $\tau_{\mu}$ could be the timescale for the gravitational settling of Helium  ($\sim 10^{9}$ yrs), applicable to near tachocline regions \citep{thoul1993element}. However, this phenomenon resembles an advection process rather than relaxation. For all these reasons, we may consider molecular diffusion to be most accurately related to $\tau_{\mu}$. {\color{black} 

Notwithstanding the choice, the small value of $\tau_F$ near the tachocline activates the nonlinear advection limit for $\mu$, implying that the local meridional flow will erase compositional gradients as seen in the simulations in the next section. Moreover, as mentioned before, the upper bound does not hold for transient/evolving systems. The Sun and similar stars, given their tiny internal viscosity, have been envisaged to be rotationally evolving over their main-sequence lifetimes \citep{banik2024meridional}. The only circumstance where the Sun might be in a rotational steady state is the addition of horizontal turbulence at the tachocline \citep{spiegel1992solar} or a magnetic field deep inside \citep{gough1998inevitability}. In that case, $\tau_F$ will be even smaller, truncating the penetration of meridional circulation close to the tachocline. An improved approach would be use a vertically varying $\tau_F$, with smaller values close to the radiative-convective boundary and larger values in deeper regions, although an exact form remains a matter of further discussion.} Given the uncertainties in $\tau_{\mu}$ and $\tau_{F}$, we leave the full calculation of the degree of penetration of meridional circulation for future work and simply focus on the self-contained fluid mechanics problem.
}

\begin{figure}
    \centering
    \includegraphics[width= 0.8\linewidth]{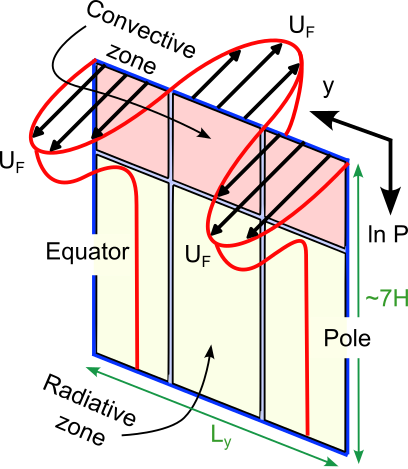}
    \caption{The two-layer setup used in our formulation shows the top and bottom layers modeling the convective and radiative zones, respectively. The forcing wind shown in red is maximum in the top layer and dies down in the bottom forced region. The blue box represents our computational domain with the origin towards the equator and the upper convective zone, with the diagonally opposite end towards the pole and the bottom of the radiative zone. {\color{black} For the solar radiative zone, the vertical extent of the domain is about 7 pressure scale heights H. The equator and pole are only representative and do not imply exact limits.}}
    \label{fig:2}
\end{figure}

 \begin{table*}[ht]
\hspace{-2.5cm}
  \caption{Numerical values used for Jupiter’s upper‐atmosphere and corresponding scaling quantities used in simulation (from \citealp{showman2006deep}). The values of $\tau_R, \tau_F$ and $\tau_{\mu}$ are two OOM smaller in the top-layer than in bottom-layer as given below. The bold-faced quantities apply to simulations in Appendix \ref{app:scale}. The rest apply to T0, T1 and T2. }
  \label{tab:jupiter_params}
  \begin{tabular}{@{}lllll@{}}
    \toprule
    Parameter & Symbol & Value & Units & Notes \\
    \midrule \midrule
    Coriolis parameter                 & $f$               & $1.8\times10^{-4}$     & rad\,s$^{-1}$                         &  \\
    Buoyancy frequency                 & $N$               & $2\times10^{-6}$       & Hz                                    &  \\
    Scale height                       & $H$               & $10^{7}$      & cm                                    &  $\mathbf{2\times10^{-4}}, \mathbf{2\times10^{-3}}$ \\
    Forcing wind speed                 & $U_F$             & $10^{4}$      & cm\,s$^{-1}$                          &  \\
    Meridional length scale            & $L_y$             & $1.26\times10^{9}$     & cm                                    &  \\
    Universal gas constant             & $\mathcal{R}_u$   & $8.32\times10^{7}$     & erg\,mol$^{-1}$\,K$^{-1}$             &  \\
    Vertical compositional gradient     & $\alpha$          & $10^{-2}$     & g\,mol$^{-1}$                         &  $(=\partial \mu/\partial \ln p)$ \\
    Mean molecular-weight     & $\mu_{\rm min}$      & $2.4$                  & g\,mol$^{-1}$                         &  \\
    Viscous/momentum forcing timescale & $\tau_F$          & $100$                  & yr                                    &  \\
    Radiative timescale                & $\tau_R$          & $10^{4}$      & yr                                    &  \\
    Compositional relaxation timescale & $\tau_\mu$     & $10^4$                  & yr                                    & assumed \\
    \midrule
    Meridional velocity scale          & $V$               & $0.017$                 & cm\,s$^{-1}$                          & from eq.~\eqref{eq:20} \\
    Vertical velocity scale            & $W$               & $1.5\times10^{-11}$ & \textit{H} s$^{-1}$                              &   \\
    Temperature perturbation scale     & $T_s$             & $5\times 10^{-5}$                   & K                                     & $\mathbf{0.05, 50}$ \\
    Molecular‐weight perturbation scale& $\mu_s$           & $0.044$     & g\,mol$^{-1}$                         &  \\
    \midrule
    Horizontal advective timescale     & $\tau_{\rm mer\_adv}$  & $2.4\times10^{3}$   & yr                                    & $L_y/V$ \\
    Vertical advective timescale       & $\tau_{\rm vert\_adv}$ & $2.4\times10^{3}$   & yr                                    & $H/W$ \\
    Zonal advective timescale          & $\tau_{\rm zon\_adv}$  & $2.18$              & yr                                    & $R/U_F$ \\
    \bottomrule
  \end{tabular}
\end{table*}

\section{Numerical simulations}\label{sec:4}


\subsection{Setup} 

Our computational domain is divided into a top driving layer resembling the momentum forcing applied by the convection zone, and a bottom driven layer resembling the passive radiative zone; see Figure \ref{fig:2}. While most simulation parameters are constant throughout the domain, some vary between the top and bottom layers. The zonal wind int the forcing and passive layers are linearly relaxed to $U_F$ and 0 cm/s. This happens over the frictional timescale $\tau_F$, which for the top layer is kept two orders of magnitude smaller than that for the bottom layer following \cite{showman2006deep}. Similarly, for the radiative and $\mu$-diffusion timescales, $\tau_R$ and $\tau_{\mu}$, respectively. This mimics the stronger diffusion due to enhanced mixing by convective currents experienced by the top layer, as compared to the bottom layer, which is stably stratified. The discontinuity between the two layers is resolved using a smoothed Heaviside step function $H(x)$ as follows, 
\begin{equation}
    H(x)=1+\tanh(Ax)
\end{equation}
where $A$ determines the steepness of the transition.

At this point, it is useful to differentiate between our setup and that of \cite{banik2024meridional}, which starts from a similar 2D framework with two layers, but eventually simplifies it to a linear-advection diffusion equation in 1D for only the bottom layer. In their work, the forcing from the convection zone is applied as a Dirichlet boundary condition at the top of the radiative zone, without accounting for the exchange of momentum between the two layers. This allows more momentum to be pumped into the bottom layer, than if the top layer were to lose some of it to the bottom through exchange.

{\color{black} The role of friction/drag here is counter-intuitive. It does not help the zonal wind spread deeper but rather prevents it. In the absence of drag, momentum from the upper layer is conveniently transferred to the lower layer via adiabatic adjustment and subsequent radiative diffusion, both of which act against the stratification \citep{spiegel1992solar,banik2024meridional}. Thus, the top layer may never reach the forced velocity $U_F$, as some of the pumped momentum is distributed vertically downwards. However, when horizontal turbulence acts \citep{spiegel1992solar}, it stops the penetration of the zonal wind, thereby trapping the momentum in the upper layer. This results in a top-layer velocity effectively closer to $U_F$ with little to no momentum in the passive bottom layer. Please refer to Appendix \ref{app:scale} for a discussion on appropriate zonal velocity scaling in both the layers.} 

\subsection{Numerical method}

We use Dedalus2 \citep{burns2020dedalus} to solve equations \eqref{eq:15}-\eqref{eq:19} in the domain. Dedalus is a pseudo-spectral nonlinear PDE solver widely used for fluid dynamics problems that allows symbolic entry of equations, making it convenient to add and drop terms. We use a Fourier basis in the meridional direction with periodic boundary conditions. This assumption is a stretch for the Sun given that it has at most one complete meridional circulation cell in the upper/lower hemisphere, the bare minimum for periodic boundaries. However, it applies better to planets like Jupiter, which have several bands of zonal flow moving in opposite directions. {\color{black}In the vertical, we use a Chebyshev basis, with impermeable boundaries ($w=0$), and Neumann conditions for $\mu$ and $T$ at the top and bottom of the domain: $\p (T, \mu)/\p z = 0$. We also try other combinations of BCs and find them to have negligible impact on the overall dynamics of the problem.} The horizontal-vertical resolution of the domain is $32 \times 128$. Dedalus permits non-constant coefficients, crucial for our two-layer formulation with varying $\tau_F, \tau_R, \tau_{\mu}$ and $u_F$. The initial conditions of the problem are set to zero for rotating-frame velocities, temperature, and molecular-weight perturbations. Post discretization, the variables are evolved through an RK222 timestepper. To determine the steady state, we calculate the total kinetic energy of the system as a sum of its zonal and meridional components.


\subsection{Results}

\begin{figure*}
    \centering
    \includegraphics[width=0.8\linewidth]{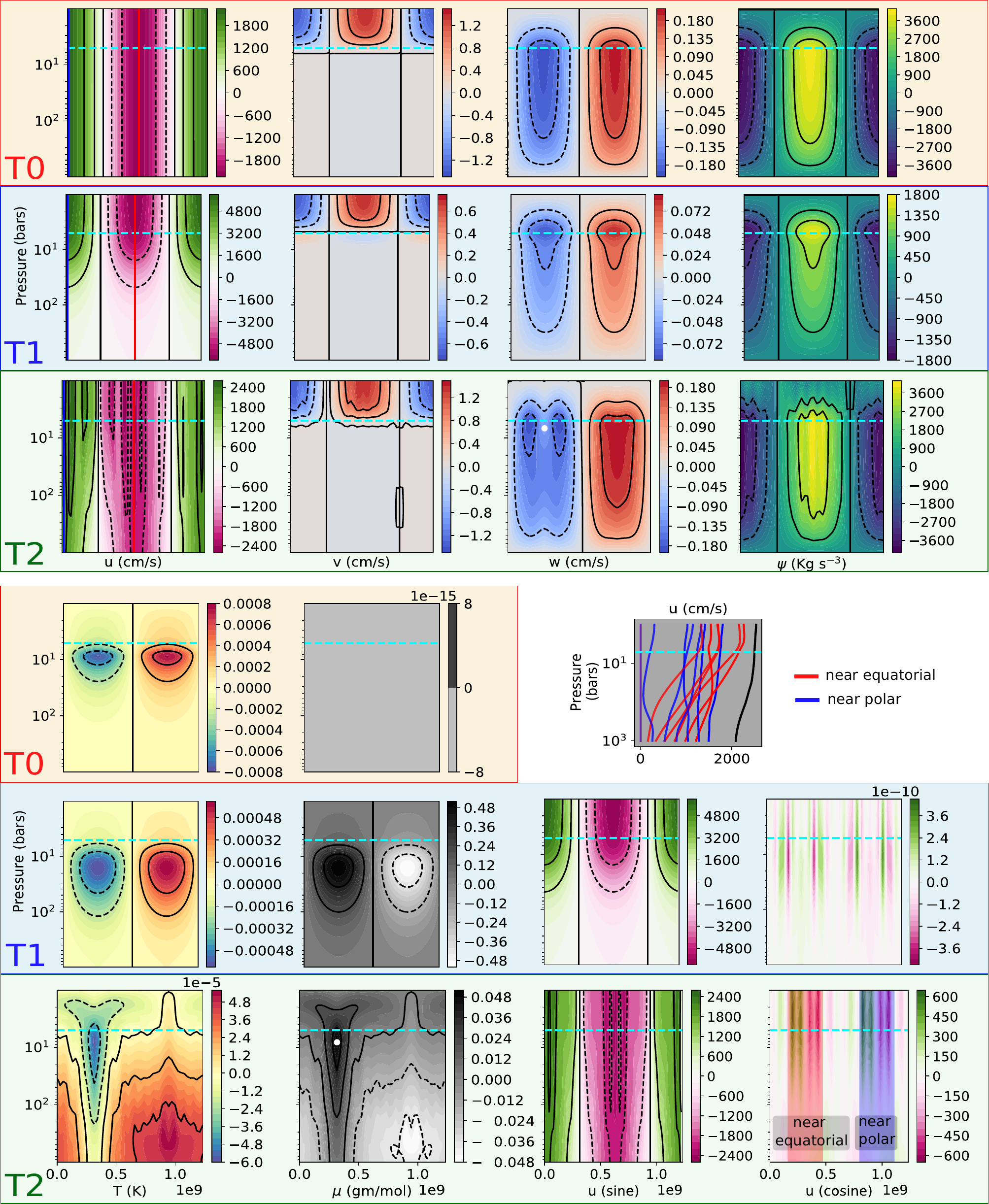}
    \caption{Contours of zonal velocity $u$, meridional velocities $v$ and $w$, temperature and molecular-weight perturbations $T'$ and $\mu'$, and the streamfunction $\psi$ at steady state for T0 (red), T1 (blue) and T2 (green) from table \ref{tab:onlytable}. The bottom two rows also show the dominant sine and cosine components when the zonal velocity field is decomposed in the horizontal/Fourier direction. The interface between the top (forced) and bottom (passive) layers is marked by the dashed cyan line. The red and blue, dashed and solid lines represent horizontal locations in the domain where profiles of respective quantities are extracted and plotted in Figure \ref{fig:4} later. T0 represents the control simulation from \cite{showman2006deep} with a Brunt frequency of $2 \times 10^{-6}$ rad/s without a background molecular-weight gradient. The zonal velocity ($u$) and the meridional circulation ($w$) are seen to penetrate all the way to the bottom. In T1, we add a background $\mu$-gradient while keeping our set of equations linear and find that it restricts the penetration of the circulation, resulting in a baroclinic structure. However, when the nonlinear terms are added back in T2, the circulation penetrates deep again. This happens because the generated $\mu$-field is mixed by the circulation, reducing the magnitude of horizontal gradients that otherwise support the truncation of penetration (compare $\mu$ fields of T1 and T2). The stand-alone floating plot shows vertical zonal velocity profiles extracted from the bottom-right panel of T2. In the nonlinear limit, an equator-pole non-uniformity is developed as the red (near-equatorial) profiles are generally more baroclinic than the blue (near-polar) ones. In conclusion, the penetration of meridional circulation is strongly reduced by the presence of $\mu$-gradients in the linear limit, while in the nonlinear limit, the role played is not as significant. }
    \label{fig:3}
\end{figure*}

{\color{black} The principal goal of this section is compare the effect of mean-molecular-weight gradients on the penetration of meridional circulation in linear and nonlinear limits. This marks an important distinction between our work and that of \cite{charbonneau1988two} and \cite{theado2003coupling}, which mostly explore linear effects. As we shall see below, this leads to major differences in the nature of the meridional flow. The names, corresponding descriptions, and parameter regimes of different simulations are given in Table \ref{tab:onlytable}. 

\subsubsection{T0 --- Reproducing \cite{showman2006deep}} \label{sec:showman}

Before proceeding with more involved simulations that include molecular-weight gradients, we verify our 2D model using existing solutions of \cite{showman2006deep}, which we also call SGL06. This is important as the current setup has never been solved numerically in 2D; SGL06 applies the `meridionally-periodic' simplification similar to our equation \eqref{eq:9}, leading to a 1D treatment, while we retain the meridional coordinate to permit variations that might arise from nonlinearities later on. All relevant physical quantities crudely fit Jupiter's upper atmosphere and are taken to be the same as in SGL06 (see Table \ref{tab:jupiter_params}). Also, the interface between the top and bottom layers is non-differentiable in SGL06, while ours applies a smooth transition. This is to prevent the build-up of energy at smaller spatial scales, leading to the Gibbs phenomenon in numerical simulations.

From Table \ref{tab:onlytable} we see that $Ro\cdot Pr$ for T0 is of order-of-magnitude 1, implying nonlinear effects to be non-negligible for the temperature perturbation. However, to match with \cite{showman2006deep}, we remove the nonlinear terms in equation \eqref{eq:16}, similar to SGL06. The set of red-background panels in Figure \ref{fig:3} shows various computed fields at steady state for $N=2\times 10^{-6}$ Hz. To compare, the reader is referred to the bottommost row in Figure 3 of SGL06. Our numerics reproduce the results in SGL06 very closely, although there are minor differences. For example, the maximum temperature perturbation is nearly 0.02 K in SGL06, while it is only around $0.0008$ K in our simulations (maximum  $T$ for T0). This is due to our relatively smooth transition between the top and bottom layers applied, which tends to diffuse strong jumps. From the first panel corresponding to zonal velocity $u$, we see a strongly barotropic structure indicating full penetration at steady state, commensurate with SGL06, their maximum quantitatively larger than ours by about 80 cm/s. However, both models are idealized, and the nature and degree of penetration match reasonably well, thereby capturing the dominant physics of the system.

In the same figure, we also show the meridional velocities $v$ and $w$. The $w$ profile is non-zero in the bottom layer, implying that the vertical component of the meridional circulation reaches depths with the corresponding zonal flow, the boundaries being impermeable ($w=0$). If the penetration was truncated, the $w$ velocity would be zero in the lower layers, as we see in T1 later. A similar pattern is seen for the streamfunction calculated from the horizontal velocity by performing a cumulative trapezoidal integral of the $v$ field in the vertical \citep{sidorenko2020simple} as 
\begin{equation}
    \psi(z)=\int_0^z v(z) dz.
\end{equation}
Comparing the magnitude and structure of the streamfunction with SGL06, we find a reasonable match. We have also run other cases with varying Brunt frequency (some cases shown in Appendix \ref{app:scale}), rotation, frictional, and radiative timescales, and observed the physics of the penetration of meridional circulation to be consistent with literature \citep{spiegel1992solar,garaud2002rotationally,gilman2004limits,garaud2008penetration}.



\subsubsection{T1 --- linear with  with $\mu$-gradient}

Now, keeping the equations linear, we introduce the background compositional stratification in T1. Linearity allows the ensuing $\mu$ and $T$ fields to still fit the sinusoidal dependence in equation \eqref{eq:8}. Note that in T1, we are in the nonlinear regime as both $Ro \cdot Pr$ and $Ro \cdot Sc$ are of order-of-magnitude 1, implying nonlinear advective terms to be significant in equations \eqref{eq:16} and \eqref{eq:17}. However, to prevent them from dominating, we externally set them to zero and check the results in the linear limit. 

Structurally, the fields look similar to T0, with the addition of a perturbation $\mu$ field generated by vertical advection of the compositional gradient, similar but opposite in signs to the $T$ distribution, commensurate with the explanation in \S 3.1 of \cite{vauclair2003coupling}. We see that the compositional stratification truncates the penetration of meridional circulation, leading to a stronger baroclinic profile compared to T0, confirming the predictions from the linearized theory. The corresponding fields are shown in the blue background panels in Figure \ref{fig:3}.

\subsubsection{T2 --- nonlinear with $\mu$-gradient}

The run T2 is the same as T1, except the nonlinear terms are now added back in equations \eqref{eq:16} and \eqref{eq:17}. From the cosine component of the zonal velocity (bottom right corner in Figure \ref{fig:3}), we see that the dominant mode is of a higher order than the lowest (driving) sine mode. This contribution comes from the nonlinear terms in the $\mu$ and $T$ equations where two lower modes interact to generate higher order modes. Specifically noteworthy is the zonal velocity field $u$, which is more baroclinic \ie has a stronger vertical gradient towards the equator than the pole. The magnitude of this non-uniformity may be noted in the cosine contribution to the zonal velocity field. To demonstrate this further, we plot vertical profiles of zonal velocity in red and blue for near-equatorial and near-polar regions, respectively, shown as the stand-alone panel in Figure \ref{fig:3}. The near-equatorial profiles show visibly stronger vertical shear than the near-polar ones, denoting non-uniform penetration of zonal velocity. An explanation for the physical mechanism driving this process is described in Appendix \ref{app:pm}.

Note that the maximum magnitudes of the $\mu$-perturbation from T1 and T2 are separated by an order of magnitude. The horizontal gradients of $\mu$ are weaker for T2, and hence are less efficient in truncating the penetration via the thermal wind relation. Also note that the steady-state field for $\mu$-perturbation has heavier material towards the top (grey-black) and lighter material at the bottom (white-light grey). The vertical gradient is thus opposite in sign to the background gradient of $\mu$, which is stably stratified, effectively overturning the stratification (net field = background + perturbation). This shows that mixing occurs in the presence of nonlinear terms. This erasure of gradients is fundamental to the meridional circulation, and the addition of horizontal turbulence may only enable further local mixing near the tachocline \citep{zahn1992circulation}, but is not necessary to explain overturning in general. 

Addressed previously in the context of Helium settling \citep{charbonneau1988two}, rotational mixing in stars due to meridional circulation has been remarked to have little impact due to nonlinearities, encouraging further studies with linear approximations \citep{charbonneau1991meridional, paper3}. Their setups are, however, different from ours; empirical prescriptions are used to obtain diffusion velocities/coefficients for specific elements, rather than solving the fully coupled hydrodynamics problem. Having catered to the latter here, we find that nonlinearities are, in fact, the sole driver of overturning of gradients in our setup.

\begin{figure}
    \includegraphics[width= 0.95\linewidth]{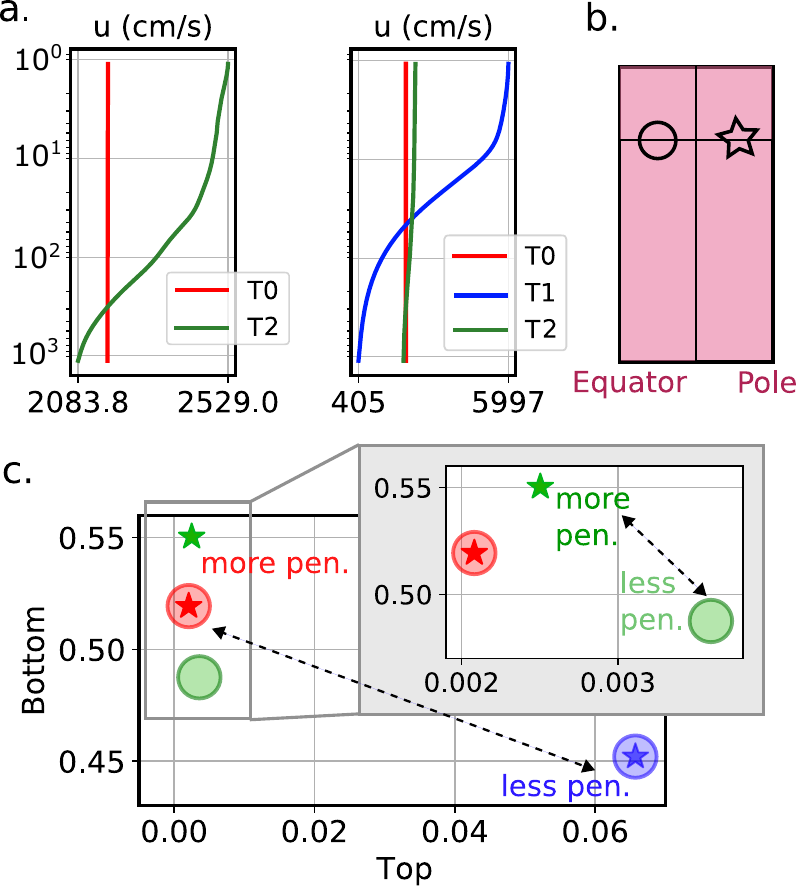}
    \caption{(a) Comparison of extracted vertical profiles of zonal velocity from the locations of blue and red vertical lines in Figure \ref{fig:3} for T0, T1, and T2 showing different degrees of burrowing of the zonal wind $u$. It is most truncated in the linear limit and in the presence of compositional gradients (T1). In the nonlinear limit (T2), truncation still occurs, but the baroclinicity in the profile is comparable to the case without $\mu$-gradients (T0). (b) Schematic showing the near-equatorial (circle) and near-polar (star) regions from our 2D computational f-plane. These markers are used in the next figure. (c) Comparison of the zonal kinetic energy density (ZKED) in near-equatorial and near-polar regions as a fraction of the total ZKED. The $x$ and $y$ axes represent ZKED in the top and bottom layers, respectively. In general, cases of greater penetration appear near the top-left corner, implying that the bottom layer has relatively more ZKED and the top layer has less. Diagonally opposite (shown with dashed line) regions represent less penetration. T0 (red) and T2 (green) are more penetrating than T2 (blue). Moreover, the markers for T0 and T1 overlap as they are linear and do not have equator-pole non-uniformity. T2, being nonlinear, shows more penetration near the poles (star) and less towards the equator (circle; shown in inset), although not very significant. }
    \label{fig:4}
\end{figure}

\subsection{Discussion and caveats}

We now compare the three simulations, T0, T1, and T2. Figure \ref{fig:4}(a) shows the absolute vertical profiles of zonal velocity extracted from the locations corresponding to the red and blue lines on the $u$ field in Figure \ref{fig:3}. These locations correspond to the maximum absolute values in the $u$ field and also overlap with each other due to the horizontal periodicity in the domain. We see that when a compositional gradient is added (T1), keeping the equations linear, it strongly truncates the penetration of zonal velocity, and therefore the corresponding meridional circulation. However, if the nonlinear terms are permitted to operate (T2), the degree of truncation is significantly restricted. This is because the nonlinear advection terms cause mixing, erasing the horizontal gradients of $\mu$ and $T$ that are responsible for the steepening of the vertical gradient of the zonal wind through the thermal wind relation. 

Figure \ref{fig:4}(c) shows the same by comparing the fraction of total zonal kinetic energy density (ZKED) over our 2D domain in the top and bottom layers. The circular and star-shaped markers represent the near-equatorial and near-polar regions as shown in Figure \ref{fig:4}(b). The markers corresponding to T1 (blue) have higher top and lower bottom layer ZKED, implying that the total momentum pumped in the top layer does not leach as efficiently into the bottom layer as for T0 (red) and T2 (green). Additionally, for T2 (green) we find that the near-equatorial region (circle) has greater top and lower bottom layer ZKED than the near-polar region (star) signifying non-uniform penetration in the domain. However, this non-uniformity is not significant for the case shown.

Given that our 2D domain is set on an f-plane similar to \cite{banik2024meridional} with periodicity in the meridional direction, and a symmetric forcing profile as shown in Figure \ref{fig:2}, the generation of the equator-pole non-uniformity is rather unexpected. More detailed simulations in spherical geometry, capturing the full variation of the Coriolis parameter \citep{korre2024penetration}, and applying a more realistic forcing profile in the top layer, could lead to further asymmetries, the effects of which are not captured here. Our model is also limited due to small resolution. Increasing resolutions in the meridional direction showed the strength of the higher-order cosine component to increase slightly. However, a more careful study remains as a scope for future work. Despite these caveats, the leading order conclusions are significant extensions of the current understanding of the effect of compositional gradients on the penetration of meridional circulation in stratified environments.


\section{Applications}\label{sec:5}

Stellar spin-down, primarily driven by the interaction between stellar winds and magnetic fields, is a crucial process in the evolution of rotating stars. Stars with convective outer layers generate magnetic fields through dynamo action, which extend into regions of escaping material or stellar winds. The stellar wind, carrying away the magnetic field lines, creates a torque on the star which penetrates through the convection zone, thereby torquing the radiative zone underneath \citep{kawaler1988angular}. This torque counters the existing rotational state, and the interior circulation adjusts by moving the necessary amount of angular momentum to the surface, effectively slowing down the core \citep{zahn1992circulation}. Here, the interior circulation refers to the combination of the zonal and meridional circulations, or the differential rotation of the radiative interior. The nature and evolution of this circulation, under the effect of molecular-weight gradients as described in the previous sections, is thus important to accurately quantify the degree of outward angular momentum transport.


\subsection{Angular momentum transport in stellar evolution models}\label{sec:5.1}

Observations suggest that compact objects like black holes and neutron stars rotate more slowly than evolutionary models predict \citep{fuller2019most}, requiring enhanced angular momentum transport in modeling stellar evolution. The current implementation of angular momentum transport via meridional circulation in state-of-the-art stellar evolution models \citep{heger2000presupernova} follows the outdated Eddington-Sweet theory \citep{busse1981eddington}. Moreover, choking of this circulation by molecular-weight gradients \citep{mestel1953rotation} originally suggested to explain the coexistence of strong differential rotation and compositional gradients in red giants, is assumed as the standard prescription for the determination of corresponding diffusion coefficients. Updates on this hydrodynamic phenomenon presented in appendix B of \cite{spiegel1992solar} disagree with previous conclusions, but are neither discussed at length nor reproducible. These updates, though acknowledged, are not implemented self-consistently during stellar evolution owing to the computational cost of solving a fourth-order hyperdiffusion equation. To this end, an analytical solution to the time-dependent differential rotation in radiative interiors approximating that of \cite{spiegel1992solar} is offered in \cite{banik2024meridional}. Stellar and planetary evolution codes like MESA \citep{paxton2010modules} and APPLE \citep{sur2024apple} may incorporate these analytical solutions. This study informs the applicability of the same in the presence of molecular-weight gradients and also attempts to resolve existing literary disagreements. 

We have found that the penetration of meridional circulation is indeed slowed down by mean-molecular-weight gradients as shown in \cite{mestel1953rotation}. In the presence of additional mechanisms like horizontal turbulence \citep{spiegel1992solar} the penetration is choked earlier/faster by $\mu$-gradients. However, the choking in the linear limit is significantly larger than in the nonlinear limit. This is because, in the latter, mixing causes erasure of the horizontal gradients in $\mu$, {\color{black} that crucially contribute to choking following the modified thermal wind balance equation.}  The degree of choking in the nonlinear limit is fairly minimal and not determined analytically. The above conclusions disagree with those reached in Appendix B of \cite{spiegel1992solar}, which says that $\mu$-gradients add to the diffusion provided by radiative effects, thereby helping the penetration of the meridional circulation. Additionally, if the nonlinear limit is true for any system, the equatorial and polar penetration is non-uniform. The extent of this nonuniformity will be contingent on a future parameter space survey.

Our theory essentially captures the spread of differential rotation profiles through stellar interiors in the presence of molecular-weight gradients due to convective forcing originating from stellar winds \citep{spiegel1992solar} rather than rotationally driven Eddington-Sweet flows \citep{mestel1953rotation}. The conclusions broadly agree with \cite{mestel1953rotation}. In this way, it captures the theoretical development of both articles using a simplified atmospheric science framework \citep{showman2006deep, haynes1991downward, banik2024meridional}. Additionally, the analytical solutions to differential rotation profiles possible from our 1D formulation in the linear limit may be incorporated in evolutionary codes to probe for magneto-/hydrodynamic instabilities, further enhancing our understanding of angular momentum transport in stellar interiors; see \cite{banik2024meridional} for exact functional forms.

\subsection{The solar tachocline}

{\color{black} For the Sun, helioseismic inversions suggest the existence of two things: the tachocline and strong vertical gradients of heavier elements beneath it; see Figure 3.4(b) of \cite{Christensen-Dalsgaard_Thompson_2007}. Reconciling the exact dynamics of the tachocline with observed abundances is an ongoing effort. Nonetheless, we present two  tentative pictures here. 

The first corresponds to a slow rotational evolution of the interior. Eddies from the  convection zone enhance the diffusivity of the tachocline region, lowering the effective viscous timescale by several orders of magnitude. These give rise to strong meridional velocities following the scaling in equation \eqref{eq:20} rendering the system nonlinear to compositional adjustment, and erasing gradients of mean molecular weight. {\color{black} This traps momentum in the upper layers allowing only a smaller fraction to make its way to the interior.} Below the tachocline, the viscous timescale increases, making the meridional advection timescale larger than the timescale for gravitational settling of heavy elements. Here, compositional gradients can co-exist with a weak, slowly evolving meridional flow decaying with depth, justifying helioseismic evidences for abundances below the tachocline.

In picture two, we consider a steady tachocline. Assuming anisotropic horizontal turbulence or magnetic effects to be strong enough to truncate the penetration of meridional circulation and render a steady state, the angular momentum pumped by the convection zone into the tachocline stays close underneath. This causes the erasure of compositional gradients by a shallow yet strong meridional circulation in the tachocline region. Below this, rigid rotation allows the settling of heavier elements forming vertical compositional gradients.

Thus, a vertical molecular-weight gradient may be supported in both cases, while the former has an additionally meridional gradient due to the thermal wind balance. Given a fixed Rossby number, the depth of observable compositional gradients will depend on the local Schmidt number, following inequality \eqref{eq:21}. However, both $\tau_{\mu}$ and $\tau_F$ are not well constrained quantities for the Sun, and should ideally be functions of radius. Here, we have avoided providing exact values trying justifying the above scenarios, and the complete investigation remains as a scope for future work.}



\section{Conclusions}\label{sec:6}

We address the problem of penetration of meridional circulation into stratified radiative stellar and planetary interiors under the influence of stable gradients of mean-molecular-weightwhen driven by zonal torques from the top, similar to \cite{spiegel1992solar}. For this, we extend the \textit{downward control principle} from atmospheric science to account for molecular-weight gradients, employing a combination of analytical and numerical methods to investigate the problem. We find the following results. 

Linear truncation: Our linearized one-dimensional molecular-weighted treatment shows that the penetration of meridional circulation is resisted by molecular-weightstratification, similar to thermal stratification, therefore slowing the burrowing down as predicted by \cite{mestel1953rotation}. However, it does not stop the penetration altogether, unless external mechanisms such as horizontal turbulence or magnetic fields are invoked. If so, the meridional circulation is unable to penetrate deeply, and a steady state is reached faster than without $\mu$-gradients. 

Transition to nonlinear regimes: The linear regime of compositional advection holds so long the product of Schmidt and Rossby numbers is much less than 1, else nonlinear effects dominate. The use of dimensionless numbers generalizes the application to stars, planets, and other systems alike. When nonlinear terms are included, gradients of molecular-weightare erased by the meridional flow because of mixing, thereby allowing penetration not possible in the linear limit. This mixing is dependent on the strength of the local meridional velocities, and additional turbulence need not be invoked separately for the purpose. With nonlinearities, an equator-pole non-uniformity is developed, with one region marginally more baroclinic (less penetration) than the other.

{\color{black} Implications for Sun-like stars: Long viscous timescales typical to the solar radiative zone suggest weak meridional flows, and hence linear advection, given the upper/forcing boundary is rigid. However, penetrative convection and invoked horizontal turbulence may lower $\tau_F$ significantly near the base of the solar convection zone, pushing it into the nonlinear limit of compositional distribution, thereby causing mixing by strong meridional currents {\color{black} trapping momentum regionally}. Beneath the tachocline where molecular effects dominate, a transition to linearity is possible. Consequently, gravitational settling of heavier elements is permitted, supporting helioseismic inferences of mean molecular-weight stratification just below the convection zone. 

Inside quiescent radiative zones in general, our 1D linearized theory provides analytical solutions to the long-term penetration of differential rotation similar to \cite{banik2024meridional}. These profiles are useful to look for fluid/magnetic instabilities that further enhance the outward transport of angular momentum in stars.

Future work must look into varying viscous/eddy and compositional diffusion appropriately to constrain the structure of the solar tachocline appropriately with observed abundances. Detailed simulations in spherical coordinates, including the variation of the Coriolis parameter and asymmetry of the forced wind may be performed, both of which could potentially contribute to additional equator-pole non-uniformity. A bigger parameter space survey based on the non-dimensional numbers relevant to the system and the specific systems that lie in those regions would be interesting to investigate. Additionally, an order-of-magnitude resistance to the penetration offered in the nonlinear limit could be explored.}

\section*{Acknowledgements}
D.B. thanks Thaddeus D. Komacek and Marta L. Bryan for the extensive discussions and feedback on the paper. D.B. is grateful to the anonymous reviewer whose comments improved the manuscript significantly. D.B. is supported by the Dept. of Physics, University of Toronto. D. B. also thanks Nicholas Brummell for related discussions on the topic. K.M. is supported by the National Science and Engineering Research Council of Canada. E.H.A acknowledges partial support from NSF grant PHY-2309135 and Simons Foundation grant (216179, LB) to the Kavli Institute for Theoretical Physics (KITP).

\section*{Data Availability}

The T2 simulation is available \href{https://github.com/Textydeep/Meridional-Circulation-Dedalus2/tree/main}{here}.

\section*{Softwares}

NumPy \citep{harris2020array}; SciPy \citep{virtanen2020scipy}; Dedalus \citep{burns2020dedalus}; GPT-4 \citep{openai2023gpt4}.

\appendix

\section{Derivation of thermal wind with compositional stratification} \label{app:tw}

On an $f$‑plane, assuming geostrophic balance in the horizontal and hydrostatic balance in the vertical, we focus on the meridional component due to axisymmetry. The vertical coordinate $z$ increases with pressure $p$.

\begin{align}
  f\,u_g &= -\dfrac{1}{\rho}\,\dfrac{\partial p}{\partial y},
  \label{eq:geostrophy} \\[1ex]
  \dfrac{\partial p}{\partial z} &= \rho\,g.
  \label{eq:hydrostatic}
\end{align}

Differentiating \eqref{eq:geostrophy} in $z$ and using \eqref{eq:hydrostatic},
\begin{align}
  f\,\frac{\partial u_g}{\partial z}
  &= -\,\frac{\partial}{\partial z}\left(\tfrac{1}{\rho}\,\partial_y p\right)
  = -\left[-\frac{1}{\rho^2} \partial_z\rho\, \partial_y p + \frac{1}{\rho}\, \partial_y(\rho g)\right] \nonumber \\
  &= \frac{1}{\rho^2}\,(\partial_z \rho)\,\partial_y p - \frac{g}{\rho}\,\partial_y \rho.
  \label{eq:tw_full}
\end{align}

On an isobaric surface where $\partial_y p = 0$, this reduces to the classical thermal wind:
\begin{equation}
  f\,\frac{\partial u_g}{\partial z} = -\,\frac{g}{\rho}\,\frac{\partial \rho}{\partial y}.
  \label{eq:tw_density}
\end{equation}

For an ideal gas,
\begin{equation}
  p = \frac{\rho\,R\,T}{\mu} \quad\Rightarrow\quad \rho = \frac{p\,\mu}{R\,T}.
  \label{eq:ideal}
\end{equation}
On constant-$p$ surfaces,
\begin{equation}
  \frac{\partial \rho}{\rho} = \frac{\partial \mu}{\mu} - \frac{\partial T}{T}
  \quad\Rightarrow\quad
  \partial_y \rho = \rho\left(\frac{1}{\mu}\,\partial_y \mu - \frac{1}{T}\,\partial_y T\right).
  \label{eq:rho_deriv}
\end{equation}

Substituting into \eqref{eq:tw_density} gives
\begin{equation}
  \frac{\partial u_g}{\partial z}
  = \frac{g}{f} \left( \frac{1}{T}\,\frac{\partial T}{\partial y} - \frac{1}{\mu}\,\frac{\partial \mu}{\partial y} \right).
  \label{eq:tw_general}
\end{equation}

Linearizing about a resting base state,
\begin{equation}
  u = u', \quad T = T_0(z) + T', \quad \mu = \mu_0(z) + \mu',
  \label{eq:linear_vars}
\end{equation}
and assuming small perturbations,
\begin{equation}
  \frac{1}{T} \approx \frac{1}{T_0}, \quad
  \frac{1}{\mu} \approx \frac{1}{\mu_0}, \quad
  \partial_y T = \partial_y T', \quad
  \partial_y \mu = \partial_y \mu',
  \label{eq:linear_approx}
\end{equation}
the linearized thermal wind becomes
\begin{equation}
  \frac{\partial u'}{\partial z}
  = \frac{g}{f} \frac{\partial}{\partial y} \left( \frac{T'}{T_0} - \frac{\mu'}{\mu_0} \right).
  \label{eq:tw_linear}
\end{equation}

We have used GPT-4 from OpenAI for this derivation \citep{openai2023gpt4}.

\section{Further comparison with Showman et. al. (2006) and an appropriate scale for zonal velocity for the 2D setup} \label{app:scale}

\begin{figure*}
    \centering
    \includegraphics[width=0.9\linewidth]{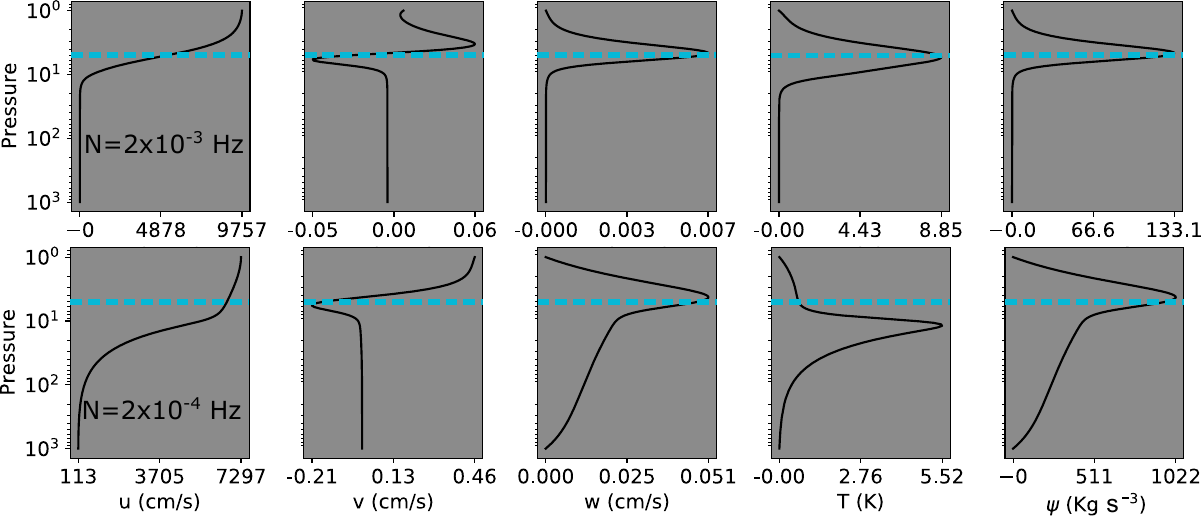}
    \caption{Vertical profiles of zonal velocity $u$, meridional velocities $v$ and $w$, the temperature perturbation $T'$ and the meridional streamfunction $\psi$ computed for Jupiter at steady state for two different Brunt frequencies $N=2\times 10^{-3}$ Hz and $2\times 10^{-4}$ Hz. The results from our 2D model compare well with literature \citep{showman2006deep}. The blue dashed horizontal line separates the top-forcing layer from the bottom-forced layer. Note that the full two-dimensional fields, which are functions of latitude and pressure, are obtained by multiplying the curves shown by a sine or cosine in latitude.}
    \label{fig:5}
\end{figure*}


For values typical to Jupiter's upper atmosphere, the scaling relations in equation \eqref{eq:20} give, $V=0.02 \text{ cm/s}, W=1.5\times 10^{-11}H\text{ /s}, T_{s}=50 \text{ K}$, and  $\mu_{s}=0.044\text{ g/mol}$ (Table \ref{tab:jupiter_params}). Note that the value of $\tau_{\mu}$ for Jupiter is not physically motivated. We may also obtain approximate horizontal and vertical advection timescales $\tau_{\text{mer\_adv}}=L_y/V=2.4\times 10^{3}\text{ yrs}$ and $\tau_{\text{vert\_adv}}=H/W=2.4\times 10^{3}\text{ yrs}$. Comparing these with the zonal advection time $\tau_{\text{zon\_adv}}=R/U_F=2.18$ yrs, we see that the timescale for meridional motion is much larger than the zonal motion. Here, $R$ is the radius of the concerned star or planet.

Comparing the values of $V, W$, and $T_s$ with the maximum magnitudes in the top row of Figure \ref{fig:5}, we see a difference in orders of magnitude. $T_s$ is larger, and $V$ and $W$ are both smaller than their corresponding simulated values. As explained earlier in \S \ref{sec:showman}, this is because of the smooth transition applied at the interface of the top and bottom layers in our 2D setup, enabling greater penetration of the zonal wind, corresponding to larger meridional velocities and smaller temperature perturbation compared to the scaling analysis. 

The scaling in equation \eqref{eq:20} predicts that $V$ and $W$ are independent of the degree of penetration. However, this is violated in our simulations. Comparing meridional velocity magnitudes in the top and bottom panels of Figure \ref{fig:5}, we see differences. This is because the zonal velocity is considered constant at $U_F$ in equation \eqref{eq:20}, while the appropriate scale for it, in the 2D setup, may be different for various degrees of penetration.


Let us call the appropriate scaling for $u$ in the bottom layer, $U_\text{scale}$. As the degree of penetration increases, so does the amount of momentum in the bottom layer, and hence $U_\text{scale}$ must also increase.  We recognize that if momentum is conserved, the total zonal momentum pumped through $U_F$ and that of the steady-state flow profile in the domain must be equal. Given $U_\text{scale}$, the functional decay of the zonal wind from the interface to the bottom layer may be obtained from the solutions provided in equation (17) of \cite{banik2024meridional}, additionally accounting for compositional gradients as in equation \eqref{eq:12.1}. By equating the vertical integrals of the forcing ($U_F$) and forced ($U_\text{scale}$) zonal momentum fields, for respective cases of penetration, we may uniquely determine $U_\text{scale}$. Thus, for the two-layer scenario, the maximum value for $U_\text{scale}$ would correspond to maximum penetration. Now, using $U_\text{scale}$ instead of $U_F$ in equation \eqref{eq:20} is reasonable, as a higher degree of penetration would imply a larger $U_\text{scale}$, and consequently larger $V$ and $W$.

\section{Discussion on Rossby number for the setup} \label{app:rossby}

It is tempting to think of $U_F/(f L_y)$ as the Rossby number of the flow. However, let us first understand what is unique about our system.

The zonal pressure gradient is zero due to the axisymmetric setup, hence Coriolis forces are not balanced by the pressure gradient \ie the flow is not in geostrophic balance. Rather, it follows from the steady-state version of equation \eqref{eq:1} that the drag term balances Coriolis forces.
Yet, we can describe the Rossby number as the ratio of inertia to Coriolis terms in equation \eqref{eq:1} as,

\begin{equation}
    Ro = \text{scale}\left( \frac{v\p u/\p y}{ fv}\right) \sim \frac{V U_F/L_y}{f V} \sim \frac{U_F}{f L_y}
\end{equation}

On the other hand, the meridional momentum equation is certainly in geostrophic balance, which leads to the thermal wind through hydrostatic balance. Similar to the above, another expression for the Rossby number would be,

\begin{equation}
    Ro = \text{scale}\left( \frac{v\p v/\p y}{ fu}\right) \sim \frac{V^2/L_y}{f U_F} \sim \frac{U_F}{f L_y}\cdot\left(\frac{V}{U_F}\right)^2
\end{equation}
 
Thus, we see that inertia terms scale differently on the horizontal plane, leading to different expressions for the Rossby number in different directions. This is because the scale for zonal velocity is $O(f \tau_F$) larger than the scale for meridional velocity from equation \eqref{eq:20}, while for most geophysical applications they are equal.

\section{Physical mechanism of non-uniform nonlinear penetration}\label{app:pm}

We provide a physical explanation of the non-uniform penetration in T2 (Figure \ref{fig:3} green panels) as follows: the background gradient advected by the meridional flow creates a perturbation $\mu$ field which is purely sinusoidal in nature, similar to the linear limit in T0 and T1, with an accumulation in the rising column and a depletion in the falling column close to the radiative-convective boundary (see corresponding $\mu$ and $T$ fields in Figure \ref{fig:3}). This field is then horizontally and vertically advected by the nonlinear terms. The meridional flow adjusts accordingly. At the peak of accumulation of $\mu$ (white dot in $\mu$ field for T2), the rising velocity (blue part of $w$ field) is reduced by the weight of the accumulated material almost creating a stagnation point around which the flow bifurcates into two faster flowing branches seen as two darker lobes on either side of the blue vertical line in Figure \ref{fig:3}. On the other side, closer to the pole, the opposite happens, the flow speeds up at the peak of depletion (red part of $w$ field). The $v$ field, directly connected to the $w$ field through the continuity equation, is consequently affected, modifying the net meridional circulation. As the flow evolves, this leads to stronger horizontal gradients of $\mu$ (checked but not shown) in the rising column compared to the falling one, leading to the non-uniform truncation of penetration following the thermal wind equation.

\bibliography{example}{}
\bibliographystyle{aasjournal}



\end{document}